\documentclass[twocolumn,amsmath,amssymb,
prl]{revtex4-2}
\usepackage{amsmath,amssymb}
\usepackage{graphicx}
\usepackage{epsfig}
\usepackage{ulem}
\usepackage[usenames]{color}

\begin{document}

\title{Nonlinear circular valley photogalvanic effect}

\author{M.~V.~Entin}
\affiliation{A.V.~Rzhanov Institute of Semiconductor Physics, Siberian Branch of the Russian Academy of Sciences, Novosibirsk 630090, Russia}

\author{V.~M.~Kovalev}
\affiliation{A.V.~Rzhanov Institute of Semiconductor Physics, Siberian Branch of the Russian Academy of Sciences, Novosibirsk 630090, Russia}

\date{\today}

\begin{abstract}
We develop a theory of circular photogalvanic effect in non-gyrotropic two-dimensional transition metal dichalcogenide
monolayers under interband optical transitions. Oblique incidence of circularly-polarized electromagnetic field or normal incidence of
elliptically polarized electromagnetic field is assumed. In contrast to the linear-in-intensity
conventional photogalvanic effect, the effect considered here arises in the second intensity order. The effect is conditioned by i)
the predominant population of the valleys by the circular in-plane electromagnetic field component and ii) the direct drift of the photo-excited
carriers by the linear-polarized in-plane electromagnetic field component in the presence of trigonal valley asymmetry.
\end{abstract}

\maketitle


\section{Introduction}

The photogalvanic effect (PGE) is the transport phenomenon consisting in the appearance of a stationary current in the sample under illumination by the external alternative electromagnetic (EM) field \cite{sturmanfridkin, ivchenko, belinicher}. This effect is not related to the light pressure, photon drag effect \cite{glazovganichev}, and non-uniformity of a sample or light intensity, like the photo-induced Dember effect or currents arising in p-n junctions under external illumination.

The photoglavanic currents appears as the second order response of charged carriers gas to the external EM field. PGE is sensitive to EM polarization and, in scientific literature, linear photogalvanic (LPGE) or circular photogalvanic (CPGE) effects are distinguished. The latter is due to the conversion of photon angular momentum to the translational motion of charge carriers. The formal phenomenological expression for CPGE reads $j_\alpha=i\lambda_{\alpha\beta}[\bold{E}\times\bold{E}^\ast]_\beta$, reflecting the photon angular momentum structure and the second-order response to the EM perturbation. Here $\textbf{E}$ is the electric field component of external EM wave.  The symmetry consideration dictates the CPGE existence in gyrotropic materials only \cite{golubivchenkospivak}.

Recently, the photo-induced transport phenomena have been actively studied in a new type of 2D systems based upon the monomolecular layers of transition metal dichalcogenides (TMD) \cite{Saito, Geim, Mak} both in normal \cite{Wang, Xiao} and superconduction \cite{Wakatsuki, Hoshino, Kovalev} regimes.  As a typical example of TMD monolayer semiconductors, we will consider below the molybdenum disulfide (MoS$_2$). This material has a $D_{3h}$ point group and its Brillouin zone consists of two nonequivalent valleys coupled by the time-reversal symmetry. From the symmetry point of view, $D_{3h}$ does not support the gyrotropy and, thus, the CPGE is forbidden in this material.

The key aim of this paper is to show that the nonlinear CPGE (nCPGE) may exist in this material and derive the corresponding theoretical description of nCPGE effect.  In contrast to the standard CPGE current, the nCPGE is the forth-order (second order, with respect to the EM intensity) response to the circular EM field and, formally, can be written as $j_{\alpha}=\chi_{\alpha\beta\gamma\delta\eta}E_\beta E^\ast_\gamma E_\delta E^\ast_\eta$. Microscopically, nCPHE arises as a forth-order response to interband optical transitions produced by the alternating EM field having the frequency that exceeds the MoS$_2$ material bandgap.

We show that an nPGE effect occurs in the circularly polarized EM field under oblique incidence to the monolayer plane. Being projected onto the monolayer plane, the circularly-polarized EM field results in the effective in-plane elliptically-polarized EM perturbation affecting the charge carriers. Formally, it can be presented as a linear superposition of two in-plane fields having circular and linear polarizations, respectively. This setup geometry has two specific advantages. The first one is that the linear component of an effective in plane EM field produces the PGE current in the system due to the trigonal intravalley symmetry of the each valley. This is known in literature underlying the intravalley PGE currents \cite{Golub1, Kovalev2}, second harmonic generation phenomenon in graphene \cite{Golub2} and valley Hall effect \cite{Golub3, Kovalev3}. At the same time, the net current vanishes due to the compensation of PGE valley currents caused by the time-reversal symmetry. The circular component of the effective in-plane EM field destroys the time-reversal symmetry and predominantly populates one of the valley resulting in the nonzero net current density in the sample. This is the second advantage of the setup considered here.

The aim of this work is to develop the theoretical description of this phenomenon. The nPGE theory requires the knowledge of the stationary, but nonequilibrium distribution function of photoexcited carriers which, in turn, requires the analysis of all relaxation processes including interband recombination, energy relaxation and intervalley scattering. Depending on the hierarchy of the corresponding times, the current magnitude may have different values.

To estimate the influence of these relaxation processes, we account them via the phenomenological relaxation times without a concrete description of the microscopic mechanisms underlying the corresponding relaxation processes. That allows one to examine the effect at different possible limiting situations.

The paper is organized as follows. In the next section we present the phenomenological description of the current from the symmetry point of view. The structure of the matrix elements describing interband optical transitions accounting for the trigonal warping of the electron dispersion in the valleys is discussed further. The next sections are devoted to the analysis of balance equations and nonequilibrium distribution functions of photoelectrons and derivation of the expression for the nPGE current density. In the final section, we discuss the results.

\section{Symmetry consideration}
\begin{figure}[t!]
\includegraphics[width=0.3\textwidth]{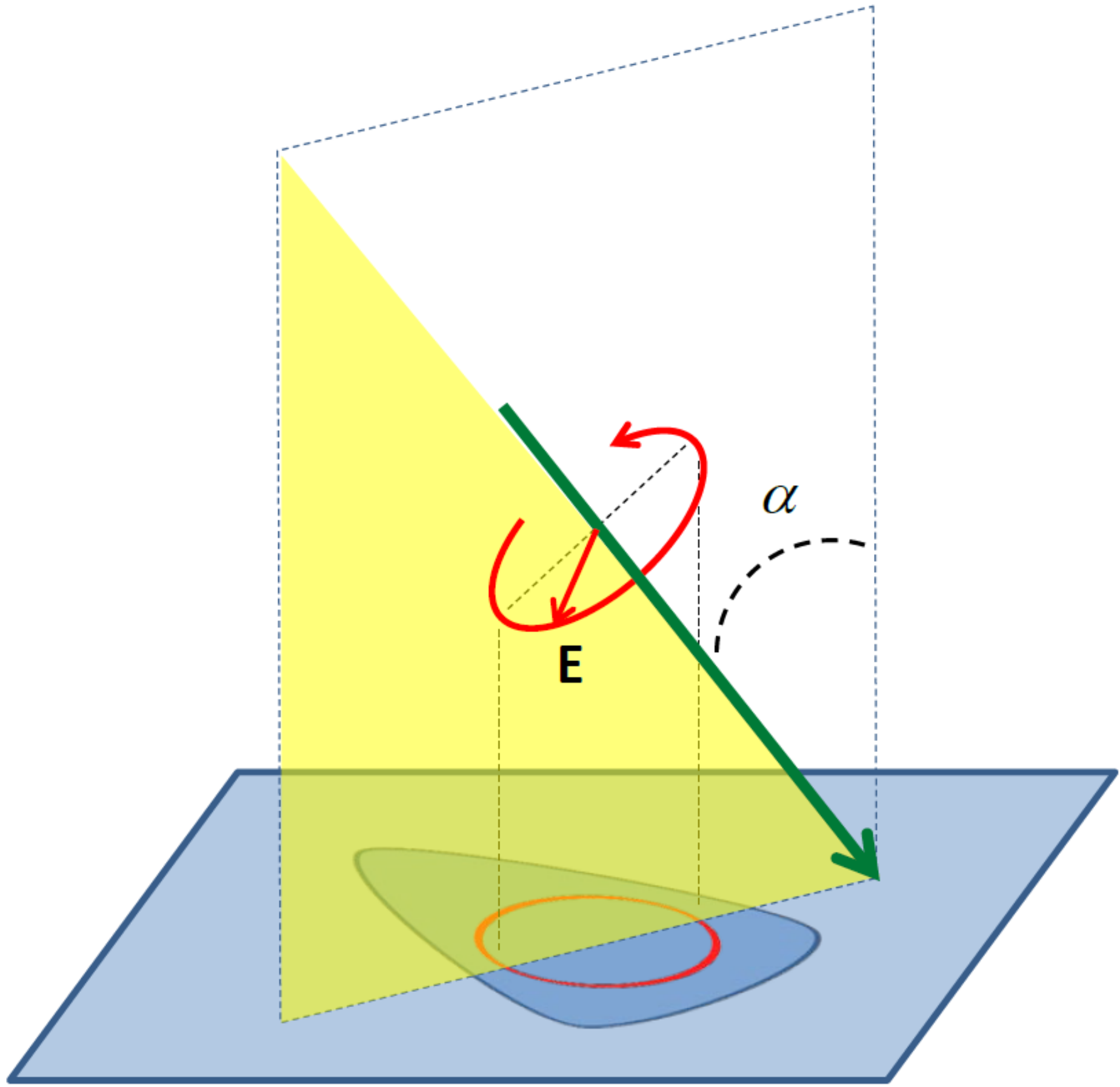}
\caption{A sketch of the system. The circular EM field under oblique
incidence produces the effective elliptically polarized in-plane
electromagnetic field arbitrarily oriented, with respect to the valley and
driving the valley population, and the in-plane carriers dynamics. }
\label{Fig1}
\end{figure}
\begin{figure}[t]
\includegraphics[width=0.3\textwidth]{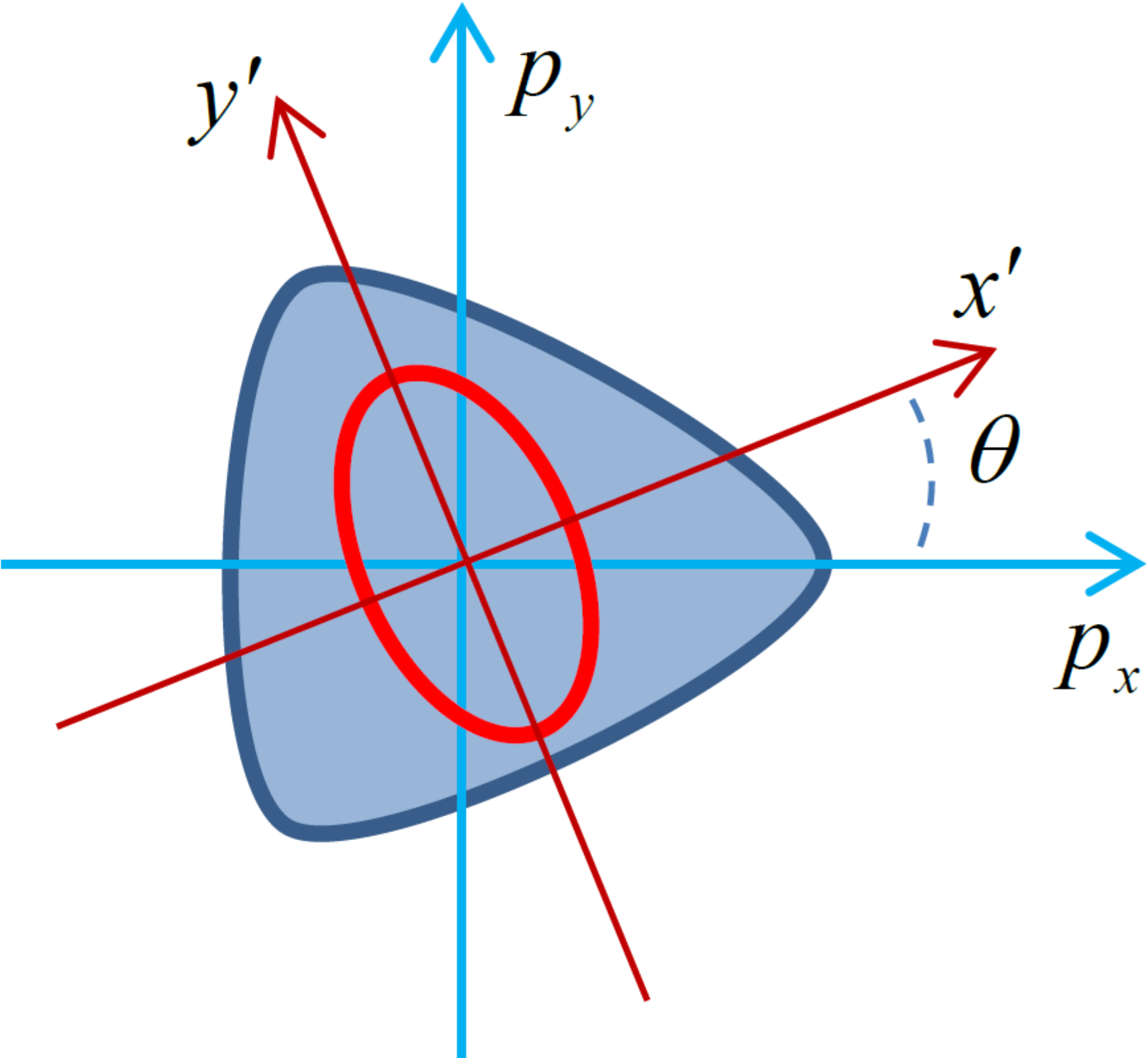}
\caption{The relative positions of the light-polarization ellipse
principal axis towards the crystallographic axis.}
\label{Fig2}
\end{figure}
Physically, as we pointed out in the Introduction, the nCPGE current occurs due to the linear component of the in-plane effective EM field. A single valley is characterized by the $C_3$ symmetry. In this case, the intravalley in-plane current density is described by a relation $j_i=\lambda_{ijk}E_iE^*_j$ with $-\lambda_{xxx}=\lambda_{xyy}=\lambda_{yxy}=\lambda_{yyx}=\lambda$. As a result, the linear PGE current is $j_x=\lambda\left(|E_x|^2-|E_y|^2\right)$ and $j_y=-\lambda\left(E_xE_y^*+E_x^*E_y\right)$, and it is characterized by a single nonzero constant coefficient $\lambda$, which is expressed via the equilibrium carriers density in the valley. In the case we consider here, the carriers density is prepared due to the selective valley photoexcitation caused by the circular component of the in-plane EM field.  Thus, the photoexcited carriers density should be proportional to the $z$-component of EM angular momentum, $\propto [\textbf{E}\times \textbf{E}^*]_z$, where the $z$-axis is directed along the monolayer plane normal. Further, we introduce the effective in-plane EM perturbation via vector potential $\textbf{A}=-i\textbf{E}/\omega$. Thus, the net nCPGE current in the sample may be phenomenologically written via the vector potential components as
\begin{gather}
\label{SymEq1}
j_x=i\chi[\textbf{A}\times \textbf{A}^*]_z\left(|A_x|^2-|A_y|^2\right),\\\nonumber
j_y=-i\chi[\textbf{A}\times \textbf{A}^*]_z\left(A_xA_y^*+A_x^*A_y\right).
\end{gather}
The expressions of Eq.\eqref{SymEq1} give the phenomenological description of nPGE effect and are also characterized by the single real parameter $\chi$.

Now consider the most important particular cases when the predicted effect can be observed. The first case is the oblique incidence of the circularly polarized EM field as shown in Fig.\ref{Fig1}.

\textit{Oblique incidence of circular field.} If the EM field has the incident components $A_0(1,i\sigma)$, then, in the in-plane valley crystallographic coordinate system, Fig.\ref{Fig2}, it reads
\begin{gather}
\label{Eq13}
A_x=A_0(\cos\alpha\cos\theta-i\sigma\sin\theta),\\\nonumber
A_y=A_0(\cos\alpha\sin\theta-i\sigma\cos\theta).
\end{gather}
In this case, the current density components have the following structure
\begin{gather}
\label{Eq14}
j_x=\chi A_0^4\sigma\cos\alpha(\cos^2\alpha-1)\cos2\theta,\\\nonumber
j_y=-\chi A_0^4\sigma\cos\alpha(\cos^2\alpha-1)\sin2\theta.
\end{gather}

\textit{Vertical incidence of elliptic field.} If the EM field has the incident components $(A_1,i\sigma A_2)$ with real amplitudes $A_1,A_2$, then, being transformed to the in-plane valley crystallographic coordinate system, it reads
\begin{gather}
\label{Eq15}
A_x=A_1\cos\theta-i\sigma A_2\sin\theta,\\\nonumber
A_y=A_1\sin\theta-i\sigma A_2\cos\theta.
\end{gather}
The corresponding current components are
\begin{gather}
\label{Eq16}
j_x=\chi\sigma A_1A_2(A_1^2-A_2^2)\cos2\theta,\\\nonumber
j_y=-\chi\sigma A_1A_2(A_1^2-A_2^2)\sin2\theta.
\end{gather}
In the further sections we derive the expression for $\chi$.


\section{Matrix elements of interband transitions}

In the lowest electron momentum order, the dichalcogenide electron Hamiltonian does  not feel the asymmetry. To include the asymmetry, one should take into account the momentum third order terms in the Hamiltonian. Hence, we use the Hamiltonian describing the band structure of the MoS$_2$ material including the valley warping and EM interaction with electrons:
\begin{gather}
\label{Eq1}
H_0=\left(
      \begin{array}{cc}
        \frac{\Delta}{2} & h_{\textbf{p}} \\
        h^*_{\textbf{p}} & -\frac{\Delta}{2} \\
      \end{array}
    \right),\,\,\,\\\nonumber
V=\left(
      \begin{array}{cc}
        0 & -evA_--2e\mu p_+A_+ \\
        -evA_+-2e\mu p_-A_- & 0 \\
      \end{array}
    \right),
\end{gather}
where $h_{\textbf{p}}=vp_-+\mu p_+^2$, $p_\pm=\eta p_x\pm ip_y$, $\eta=\pm1$ is the valley index, $\mu$ is the warping constant and $v$ is the band parameter having the velocity dimension. The interband optical transitions occur at the EM field described by the vector potential having in-plane components $\textbf{A}=(A_x,A_y)$ corresponding to the circular EM field, Eq.(\ref{Eq13}), and elliptic EM field, Eq.(\ref{Eq15}), respectively. We also use $A_{\pm}=\eta A_x\pm iA_y$ for a short-hand notation.

The bare band Hamiltonian $H_0$ has the eigenstates corresponding to valence and conduction bands
\begin{gather}
\label{Eq2}
\psi_c(\textbf{r})=\left(
                     \begin{array}{c}
                       \cos\left(\frac{\theta}{2}\right) \\
                       \sin\left(\frac{\theta}{2}\right)\frac{h^*_{\textbf{p}}}{|h_\textbf{p}|} \\
                     \end{array}
                   \right)\frac{e^{i\textbf{pr}}}{\sqrt{S}},\\\nonumber
\psi_v(\textbf{r})=\left(
                     \begin{array}{c}
                       \sin\left(\frac{\theta}{2}\right) \\
                       -\cos\left(\frac{\theta}{2}\right)\frac{h^*_{\textbf{p}}}{|h_\textbf{p}|} \\
                     \end{array}
                   \right)\frac{e^{i\textbf{pr}}}{\sqrt{S}},
\end{gather}
where $\cos\theta=\Delta/2E_c$, $\sin\theta=|h_\textbf{p}|/E_c$ and $S$ is the sample area. The electron energies in conduction and valence bands include the warping correction
\begin{gather}
\label{Eq2.1}
E_{c,v}=\pm\sqrt{\Delta^2/4+|h_\textbf{p}|^2}=\\\nonumber
\pm\sqrt{\Delta^2/4+v^2p^2+2v\eta\mu(p_x^3-3p_xp_y^2)+\mu^2p^4}.
\end{gather}
In a vicinity of conduction band bottom and valence band top, the spectrum can be simplified as
\begin{gather}
\label{Eq2.2}
E_{c,v}\approx\pm\frac{\Delta}{2}\pm\epsilon_p\pm\eta W(p_x^3-3p_xp_y^2),
\end{gather}
where $\epsilon_p=p^2/2m$, $(2m)^{-1}=v^2/\Delta$ is an effective mass and $W=v\mu/\Delta=\mu/2mv$ is a warping amplitude. Within the range of our symmetric two-band model of Eq.\eqref{Eq1}, the warping amplitudes in the conduction band and a valence band are distinguished only by a sign $\pm W$. In a more extended model \cite{falko, falko2}, these amplitudes have also different absolute values which take into account other bands. In those cases further, where this difference will play a role, we will designate them as $W_c$ and $W_v$, respectively.

An interband matrix element
$M_{cv}(\textbf{p})=M_0(\textbf{p})+M_\mu(\textbf{p})$ includes the isotropic part $M_0(\textbf{p})$ and the part due to the valley warping $M_\mu(\textbf{p})$,  where
\begin{gather}
\label{Eq3}
M_0(\textbf{p})=evA_-\cos^2\left(\frac{\theta}{2}\right)\frac{h^*_{\textbf{p}}}{|h_\textbf{p}|}-\\\nonumber
-evA_+\sin^2\left(\frac{\theta}{2}\right)\frac{h_{\textbf{p}}}{|h_\textbf{p}|},\\\nonumber
M_\mu(\textbf{p})=2e\mu p_+A_+\cos^2\left(\frac{\theta}{2}\right)\frac{h^*_{\textbf{p}}}{|h_\textbf{p}|}-\\\nonumber
-2e\mu p_-A_-\sin^2\left(\frac{\theta}{2}\right)\frac{h_{\textbf{p}}}{|h_\textbf{p}|}.
\end{gather}
The second term here,  $M_\mu(\textbf{p})$, has valley warping smallness. Assuming also $vp\ll\Delta$, the transition rate can be simplified and it can be written in the form
$|M_{cv}(\textbf{p})|^2\approx |M_{0}(0)|^2+2\textmd{Re}\,\{M_{0}(\textbf{p})M_\mu^*(\textbf{p})\}$, where
\begin{gather}
\label{Eq4}
|M_{0}(0)|^2\approx e^2v^2|A_-|^2,\\\nonumber
|A_-|^2=(\textbf{A}\cdot \textbf{A}^*)+i\eta[\textbf{A}\times\textbf{A}^*]_z
\end{gather}
and
\begin{gather}
\label{Eq5}
2\textmd{Re}\,\{M_{0}(\textbf{p})M_\mu^*(\textbf{p})\}\approx 4e^2v\mu \textmd{Re}\left[p_-A_-A_+^*\right]-\\\nonumber
-e^2v^3\mu \frac{\sin^2\theta}{|h_\textbf{p}|^2}\textmd{Re}\left[p_-^*|A_-|^2f_\textbf{p}^{*2}+p_+^*|A_+|^2f_\textbf{p}^{2}\right].
\end{gather}
The second term in Eq.\eqref{Eq5} has an additional smallness $(vp/\Delta)^2\ll1$, in comparison with the first one, and can be omitted. Thus, finally, one finds
\begin{gather}
\label{Eq6}
2\textmd{Re}\,\{M_{0}(\textbf{p})M_\mu^*(\textbf{p})\}=\\\nonumber
=4e^2v\mu\eta \left[p_x(|A_x|^2-|A_y|^2)-p_y(A_xA_y^*+A_x^*A_y)\right].
\end{gather}

\section{Balance equations and photoinduced distribution function}
In a previous section we analyzed the structure of the interband matrix elements describing the interband transitions under the EM field.
We assume that, in the equilibrium, the valence band is filled, whereas the conductivity band is empty, and the EM field producing interband transitions populates the conductivity bands. The steady-state distribution functions of photoexcited electrons in the "Left", $\eta=1$, and in the "Right", $\eta=-1$, valleys satisfy the system of balance equations in the form
\begin{gather}\nonumber
\nonumber
\frac{f_L}{\tau_r}+\frac{f_L-f_R}{\tau_v}+\frac{f_L-\langle f_L\rangle}{\tau_p}+\frac{f_L-f^0_L}{\tau_\varepsilon}+\frac{f^0_L-f^0_R}{\tau_v}=g_L,\\
\frac{f_R}{\tau_r}+\frac{f_R-f_L}{\tau_v}+\frac{f_R-\langle f_R\rangle}{\tau_p}+\frac{f_R-f^0_R}{\tau_\varepsilon}+\frac{f^0_R-f^0_L}{\tau_v}=g_R.
\label{Eq7Balance}
\end{gather}
Here $f_{L,R}$ are distribution functions of photoelectrons in the left/right valley, $\langle f_{L,R}\rangle$ are the corresponding distribution functions averaged over the isoenergetic line; $f^0_{L,R}$ describe the quasiequilibrium distribution functions of photoelectrons in a given valley, and $\tau_r,\tau_p,\tau_v,\tau_\varepsilon$ are the recombination, momentum, intervalley and energy relaxation times, respectively.  The generation rates are $g_{L,R}=2\pi|M_{cv}(\textbf{p})|^2_{L,R}\delta(E_c-E_v-\hbar\omega)/\hbar$.

The general solution of balance equations is cumbersome. Thus, we apply the following relaxation times hierarchy: $\tau_p\ll\tau_\varepsilon\ll\tau_0$ or $\tau_p\ll\tau_0\ll\tau_\varepsilon$, where $\tau_0^{-1}=\tau_r^{-1}+\tau_v^{-1}$. These inequalities correspond to slow $\tau_\varepsilon\gg\tau_0$ and fast $\tau_\varepsilon\ll\tau_0$ energy relaxation processes in comparison with the interband recombination and intervalley relaxation.  The momentum relaxation process, being the fastest processes in the system, results in the fast isotropization of the photoelectron momenta in the direction over the isoenergetic line. Average distributions $\langle f_{L,R}\rangle$ can be found from Eq.\eqref{Eq7Balance} after averaging balance equations
\begin{gather}
\nonumber
\frac{\langle f_L\rangle}{\tau_r}+\frac{\langle f_L\rangle-\langle f_R\rangle}{\tau_v}+\frac{\langle f_L\rangle-f^0_L}{\tau_\varepsilon}+\frac{f^0_L-f^0_R}{\tau_v}=\langle g_L\rangle,\\
\frac{\langle f_R\rangle}{\tau_r}+\frac{\langle f_R\rangle-\langle f_L\rangle}{\tau_v}+\frac{\langle f_R\rangle-f^0_R}{\tau_\varepsilon}+\frac{f^0_R-f^0_L}{\tau_v}=\langle g_R\rangle,
\label{Eq7BalanceIso}
\end{gather}
where the angular brackets mean averaging over the isoenergetic line as
\begin{gather}
\label{averaging}
\langle X_{L,R}\rangle=\frac{\sum_{\mathbf{p}'}X_{L,R}(\mathbf{p}')\delta(E_{c\textbf{p}}-E_{c\textbf{p}'})}{\sum_{\mathbf{p}'}\delta(E_{c\textbf{p}}-E_{c\textbf{p}'})}.
\end{gather}
In case of the isotropic spectrum, when the warping correction is neglected in Eq.\eqref{Eq2.2}, Eq.\eqref{averaging} gives the standard averaging over the momentum directions.

Let us now analyse possible solutions of Eq.\eqref{Eq7BalanceIso} for fast and slow energy relaxation processes.

\subsection{Fast intravalley energy relaxation}
If energy relaxation processes dominate over the intervalley and recombination ones, the photoelectrons lose their energy, and it leads to the formation of quasi-equilibrium distribution functions $f^0_{L,R}$ in the valleys. Due to inequality $\tau_\varepsilon\ll\tau_0$, one may leave only the third terms in Eq.\eqref{Eq7BalanceIso} and, disregarding the warping of energy spectrum, one finds $\langle f_{L,R}\rangle=f^0_{L,R}$.

We let functions $f^0_{L,R}$ have the form of quasiequilibrium Maxwell distributions $f^0_{L,R}=C_{L,R}\exp[-\epsilon_p/T]$. Normalization parameters $C_{L,R}$ can be found as follows. Integrating over momentum $\textbf{p}$ in Eqs.\eqref{Eq7BalanceIso} and taking into account that the total photoinduced electron densities in the valleys are given by
\begin{gather}
\label{Eq7BalanceIso1}
n_{L,R}=\int \frac{d\textbf{p}}{(2\pi\hbar)^2} f^0_{L,R}=C_{L,R}\frac{mT}{2\pi\hbar^2},
\end{gather}
one finds the system of equations determining their values
\begin{gather}
\label{Eq7densities}
\frac{n_L}{\tau_0}-\frac{n_R}{\tau_v}=\overline{g_L},\\\nonumber
\frac{n_R}{\tau_0}-\frac{n_L}{\tau_v}=\overline{g_R},
\end{gather}
which have the solution
\begin{gather}
\label{Eq7densitiesSol}
n_L=\frac{\tau_0^2\tau_v^2}{\tau_v^2-\tau_0^2}\left(\frac{\overline{g_L}}{\tau_0}+\frac{\overline{g_R}}{\tau_v}\right),\\\nonumber
n_R=\frac{\tau_0^2\tau_v^2}{\tau_v^2-\tau_0^2}\left(\frac{\overline{g_R}}{\tau_0}+\frac{\overline{g_L}}{\tau_v}\right),
\end{gather}
with
$$\overline{g_{L,R}}=\int \frac{d\textbf{p}}{(2\pi\hbar)^2}g_{L,R}.$$
Disregarding here the warping corrections to the matrix element and electron band energies, one finds
$$\overline{g_{L,R}}=\frac{m}{2\hbar^3}|M_{0}(0)|^2_{L,R}\theta(\hbar\omega-\Delta).$$

Finally, the normalization constants $C_{L,R}$ may be expressed via photo-induced electron densities $n_{L,R}$ using the relations of Eq.\eqref{Eq7BalanceIso1}. Here we assume that the quasiequilibrium functions are of Maxwellian type. This approach can be generalized to the distribution functions of the Fermi-Dirac form with quasiequilibrium Fermi  energies.

\subsection{Slow intravalley energy relaxation}

In the opposite limit, when $\tau_\varepsilon\gg\tau_0$, the stationary distribution is set by the time $\tau_0$, and the quasiequlibrium distribution given by $f^0_{R,L}$ is not formed. Under these conditions, one has  $\tau_\varepsilon\rightarrow\infty$, $f^0_{R,L}=0$, and the balance equations for averaged functions are reduced to the following system of equations
\begin{gather}
\label{Eq7distrEqs}
\frac{\langle f_L\rangle}{\tau_0}-\frac{\langle f_R\rangle}{\tau_v}=\langle g_L\rangle,\\\nonumber
\frac{\langle f_R\rangle}{\tau_0}-\frac{\langle f_L\rangle}{\tau_v}=\langle g_R\rangle,
\end{gather}
which have the solutions
\begin{gather}
\label{Eq7distrSol}
\langle f_L\rangle=\frac{\tau_0^2\tau_v^2}{\tau_v^2-\tau_0^2}\left(\frac{\langle g_L\rangle}{\tau_0}+\frac{\langle g_R\rangle}{\tau_v}\right),\\\nonumber
\langle f_R\rangle=\frac{\tau_0^2\tau_v^2}{\tau_v^2-\tau_0^2}\left(\frac{\langle g_R\rangle}{\tau_0}+\frac{\langle g_L\rangle}{\tau_v}\right),
\end{gather}
where
\begin{gather}
\label{Eq7distrSol2}
\langle g_{L,R}\rangle=\frac{2\pi}{\hbar}|M_{0}(0)|^2_{L,R}\langle\delta(E_c-E_v-\hbar\omega)\rangle.
\end{gather}
Thus, in contrast to the fast energy relaxation limit, the functions of Eqs.\eqref{Eq7distrSol} correspond to a very narrow photo-electron energy distribution.

\section{Photoinduced current density}
The photoinduced current density, due to interband transitions in a given $\eta-$valley, reads
\begin{gather}
\label{Eq8}
\textbf{j}^{(\eta)}=\frac{2\pi e}{\hbar}\int\frac{d\textbf{p}}{(2\pi\hbar)^2}[\tau^c_{\textbf{p}}\textbf{v}^c_{\textbf{p}}-\tau^v_{\textbf{p}}\textbf{v}^v_{\textbf{p}}]\times\\\nonumber
\times|M_{cv}(\textbf{p})|^2_{\eta}[f^v_{\eta}(\textbf{p})-f^c_{\eta}(\textbf{p})]\delta(E_c-E_v-\hbar\omega),
\end{gather}
where $f^{c,v}_{\eta}(\textbf{p})$ are the distribution functions of photocarriers in the corresponding band and valley, $\tau_{\textbf{p}}, \textbf{v}_{\textbf{p}}$ are momentum relaxation time and particle velocity in the corresponding bands. Taking into account that, for symmetric two-band model  $\textbf{v}_c(\textbf{p})=-\textbf{v}_v(\textbf{p})\equiv\textbf{v}=\textbf{p}/m$, one finds
\begin{gather}
\label{Eq9}
j^{(\eta)}_\alpha=\frac{4\pi e\tau}{\hbar}\int\frac{v_\alpha d\textbf{p}}{(2\pi\hbar)^2}|M_{cv}(\textbf{p})|^2_{\eta}\times\\\nonumber
\times[f^v_{\eta}(\textbf{p})-f^c_{\eta}(\textbf{p})]\delta(2\epsilon_p-\hbar\omega+\Delta),
\end{gather}
where we approximate the momentum relaxation times by a constant value $\tau^c_{\textbf{p}}=\tau^v_{\textbf{p}}=\tau$.

We consider the model where the valence band is filled, whereas the conductivity band is empty, and it corresponds to the undoped monolayer in the equilibrium. Due to the charge conservation under interband transitions, the nonequilibrium functions are $f^c_{L,R}=\langle f_{L,R}\rangle$ and $f^v_{L,R}=1-\langle f_{L,R}\rangle$, where $\langle f_{L,R}\rangle$ are given by either $\langle f_{L,R}\rangle=f^0_{L,R}$ for the fast intravalley energy relaxation regime, or Eq.\eqref{Eq7distrSol} in the case of slow energy relaxation.

\subsection{Current density in a fast energy relaxation regime}

The distribution function of photoelectrons in the fast energy relaxation regime, $\tau_\varepsilon\ll\tau_0$, is given by the expression
\begin{gather}
\nonumber
\langle f_{L,R}\rangle\equiv f^0_{L,R}
=\frac{\tau_0^2\tau_v^2}{\tau_v^2-\tau_0^2}\left(\frac{|M_0(0)|^2_{L,R}}{\tau_0}+\frac{|M_0(0)|^2_{R,L}}{\tau_v}\right)\times\\
\times\frac{\pi}{T\hbar}e^{-\epsilon_p/T}\theta(\hbar\omega-\Delta)
\label{Eq10}
\end{gather}
This distribution function depends on the photoelectron energy, $\epsilon_p$, and does not depend on the direction of electron momentum $\textbf{p}$. It means the finite current density occurs due to the anisotropic corrections to the interband matrix element entering Eq.\eqref{Eq9} and is given by Eq.\eqref{Eq6}.

Direct analysis shows that the net current density, $\textbf{j}=\textbf{j}_L+\textbf{j}_R$ satisfies the phenomenological expressions, Eq.\eqref{SymEq1}, with the parameter $\chi$ having the form
\begin{gather}
\label{Eq11}
\chi^{fast}=-e\pi\left(\frac{2ev}{\hbar}\right)^4\frac{\tau\tau_r\tau_v}{2\tau_r+\tau_v}\times\\\nonumber
\times
\frac{m\mu}{v}\frac{\hbar\omega-\Delta}{2T}e^{-\frac{\hbar\omega-\Delta}{2T}}\theta[\hbar\omega-\Delta].
\end{gather}

\subsection{Current density in a slow energy relaxation regime}

The distribution function of photoelectrons in a slow energy relaxation regime, $\tau_\varepsilon\gg\tau_0$, is given by the expression
\begin{gather}
\nonumber
\langle f_{L,R}\rangle
=\frac{\tau_0^2\tau_v^2}{\tau_v^2-\tau_0^2}\left(\frac{|M_0(0)|^2_{L,R}}{\tau_0}+\frac{|M_0(0)|^2_{R,L}}{\tau_v}\right)\times\\
\times\frac{2\pi}{\hbar}\langle\delta(E_c-E_v-\hbar\omega)\rangle
\label{Eqslow1}
\end{gather}
If one neglects the warping correction to the electron spectrum here, one finds $\langle\delta(E_c-E_v-\hbar\omega)\rangle=\delta(2\epsilon_p-\hbar\omega+\Delta)$. At a monochromatic excitation in
the absence of spectrum warping, the distribution function of Eq.\eqref{Eqslow1} correction
caused by the circular-polarized light is proportional to
$\delta(2\epsilon_p-\hbar\omega+\Delta)$ corresponding to a very narrow photoelectron energy distribution. When one calculates the current density in Eq.\eqref{Eq9},  this delta-function is multiplied by the same
delta-function. That leads to $\delta^2(2\epsilon_p-\hbar\omega+\Delta)$ in the current density expression. Such contribution needs to be regularized by
some widening mechanism. One of such mechanisms is the energy
uncertainty caused by the relaxation itself.

On the other hand, the energy spectrum warping results in the spread of the
photoexcited carriers energy and, simultaneously, the elastic scattering smears the carriers at the isoenergetic
line. Generally speaking, since isoenergetic line for electrons does
not coincide with that for holes, the elastic scattering would smear the carriers energy. This leads to the
liquidation of delta-squared resonance and the finite result even if the quantum
widening is taken into account.

Mathematically, it is expressed as follows. Depending on the relation between the warping correction given by $Wp_0^3$, where $p_0=\sqrt{m(\hbar\omega-\Delta)}$
and the momentum relaxation time $\tau$, we distinguish two limiting cases as $Wp_0^3\tau\ll1$ and $Wp_0^3\tau\gg1$.
%
%
%
%
%
%
%
%

In the first case, it is possible to disregard the warping corrections to the electron valley energy spectrum and use the regularization $\delta^2(2\epsilon_p-\hbar\omega+\Delta)=\frac{\tau}{2\hbar}\delta(2\epsilon_p-\hbar\omega+\Delta)$. A direct computation of the integrals in Eq.\eqref{Eq9}, gives the following expressions for parameter $\chi$ at $Wp_0^3\tau\ll1$
\begin{gather}
\label{Eqslow3}
\chi^{slow}_{\tau}=-e\pi\left(\frac{2ev}{\hbar}\right)^4\frac{\tau\tau_r\tau_v}{2\tau_r+\tau_v}\times\\\nonumber
\times
\frac{m\mu}{v}\frac{(\hbar\omega-\Delta)\tau}{2\hbar}\theta[\hbar\omega-\Delta],
\end{gather}
In the second case, $Wp_0^3\tau\gg1$, we use the current expression Eq.\eqref{Eq9} where the warping is absent, but the averaging in Eq.\eqref{Eqslow1} should be done as that in Eq.\eqref{averaging} accounting for the warping terms in energies $E_c$ and $E_v$.
Cumbersome but straightforward calculations yield
\begin{gather}
\label{Eqslow4}
\chi^{slow}_{W}=-e\pi\left(\frac{2ev}{\hbar}\right)^4\frac{\tau\tau_r\tau_v}{2\tau_r+\tau_v}\times\\\nonumber
\times \frac{2\mu\theta[\hbar\omega-\Delta]}{v|W_c-W_v|\sqrt{\pi^2m(\hbar\omega-\Delta})}.
\end{gather}

\subsection{Current density in a double-chromatic excitation regime}

In the previous sections we analyzed the nPGE appearing under the action of monochromatic illumination. There is another possibility for the appearance of the nPGE current in the sample if the latter is illumianted by two EM waves having distinguished frequencies. Let one of these EM waves have the linear polarization with frequency $\omega_1$, and the other is circularly-polarized with frequency $\omega_2$. The nPGE current density is produced if the difference of EM field frequencies is less than that of momentum relaxation time,
$1/\tau$. Otherwise, the square of the delta functions should be
replaced by their product. If so, instead of delta-functions, one can
use some of their representation, say, the Lorentzians.
The integration of two Lorentzians yields
\begin{gather}
\label{Eq12.2}
j\propto\delta(\omega_1-\omega_2)\Rightarrow\frac{1}{1+(\omega_1-\omega_2)^2\tau^2}
\end{gather}
This expression shows that the photocurrent has a resonant character at
$\omega_1\to\omega_2.$


\section{Discussion}

Now compare the current densities found in different regimes. The comparison of expressions Eq.(\ref{Eq11}) and Eq.(\ref{Eqslow3}) yields
\begin{gather}
\label{Eq12.1}
\frac{\chi^{fast}}{\chi^{slow}_\tau}=\frac{\hbar}{T\tau}e^{-\frac{\hbar\omega-\Delta}{2T}},
\end{gather}
from which we conclude that the nPGE effect has a more pronounced value in case of a slow intravalley energy relaxation process, $\chi^{slow}\gg\chi^{fast}$, because $T\tau/\hbar\gg1$. Equality $T\tau/\hbar\gg1$ reflects the fact of weak electron-impurity scattering processes for a non-degenerate electron gas.

At the same time, the comparison of Eq.(\ref{Eqslow3}) and Eq.(\ref{Eqslow4}) shows that the first expression is suppressed at the threshold absorption, $\hbar\omega\rightarrow\Delta$, whereas the latter expression, Eq.(\ref{Eqslow4}), demonstrates a sufficient increase, $j\propto(\hbar\omega-\Delta)^{-1/2}$, of the nPGE current density at the threshold.


\section{Conclusion}

We have developed the theoretical description of the novel nonlinear PGE effect in two-dimensional non-gyrotropic systems under the external uniform field affecting mobile charge carriers. It is shown that the nonlinear PGE effect occurs due to the forth-order response, with respect to the EM field amplitude. We demonstrate that the nPGE current may appear under the elliptically-polarized vertical or a circularly-polarized oblique incidence of external EM field. We have found the photoinduced electron distribution functions and estimated the current density values for the nonlinear PGE effect. It has been shown that the sensitivity of nonlinear PGE to the speed of the energy relaxation processes in photoexcited systems, in comparison with the recombination and intervalley scattering processes decreasing the material valley polarization. Our results show that the nonlinear PGE dominates in the case of slow energy relaxation processes, as compared with recombination ones.

Note that the valley current under the interband illumination of
linear-polarized light occurs in the Born approximation, without
accounting for the electron-hole interaction. This is opposite to the
case of transitions in the semiconductor with central valleys, where the
time reversibility leads to the equality of the transition probabilities
in the states with momenta ${\bf p}$ and $-{\bf p}$, thus, yielding the
spacial reflection. In the present case, the states near the valley center
are not connected by the time reversibility. That is why we did not need taking into account
the electron-hole interaction. Note also that the
latter can affect the transition rate, but its influence is weak if the
Coulomb energy is less than the excitation energy.

\section{Acknowledgement}
This paper was financially supported by the Russian Science Foundation (Project No.~17-12-01039).

\end{document}